\documentstyle[11pt]{article}
\begin{document}
\setlength{\baselineskip}{15pt}
\title{Characterization and global analysis \\ of a family of Poisson structures}
\author{Benito Hern\'{a}ndez--Bermejo $^1$}
\date{}

\maketitle

\begin{flushleft}
\noindent{\em Escuela Superior de Ciencias Experimentales y Tecnolog\'{\i}a. Edificio Departamental II.} \\
\noindent{\em Universidad Rey Juan Carlos. Calle Tulip\'{a}n S/N. 
28933--M\'{o}stoles--Madrid. Spain.} 
\end{flushleft}

\mbox{}

\mbox{}

\mbox{}

\begin{center} 
{\bf Abstract}
\end{center}
\noindent
A three-dimensional family of solutions of the Jacobi equations for Poisson systems is characterized. In spite of its general form it is possible the explicit and global determination of its main features, such as the symplectic structure and the construction of the Darboux canonical form. Examples are given.

\mbox{}

\mbox{}

\mbox{}

\mbox{}

\noindent {\bf PACS codes:} 45.20.-d, 45.20.Jj, 02.30.Jr.


\mbox{}

\noindent {\bf Keywords:} Finite-dimensional Poisson systems --- Jacobi identities --- 
3-d systems --- PDEs.

\vfill

\noindent $^1$ Telephone: (+34) 91 488 73 91. Fax: (+34) 91 488 73 38. \newline 
\mbox{} \hspace{0.05cm} E-mail: {\tt benito.hernandez@urjc.es }

\pagebreak
\begin{flushleft}
{\bf 1. Introduction}
\end{flushleft}

Finite-dimensional Poisson structures (see \cite{olv1} and references therein) have a significant presence in all fields of mathematical physics, such as mechanics \cite{haa1}, 
electromagnetism \cite{7}, plasma physics \cite{25}, optics \cite{17,dht3}, dynamical systems theory \cite{9}-\cite{28}, etc. Describing a given physical system in terms of a Poisson 
structure allows the obtainment of a wide range of information which may be in the form of perturbative solutions \cite{cyl1}, invariants \cite{byv3}, stability analysis \cite{hyct}, bifurcation properties and characterization of chaotic behaviour 
\cite{dht3}, efficient numerical integration \cite{mac1} or integrability results \cite{mag1}, 
to cite a sample. 

Mathematically, a finite-dimensional dynamical system defined on $I \! \! R^n$ is said to have a Poisson structure if it can be written in terms of a set of ODEs of the form: 
\begin{equation}
    \label{plnnham}
    \frac{\mbox{d}x_i}{{\mbox{d}t}} = \sum_{j=1}^n J_{ij} \partial _j H 
	\; , \;\:\; i = 1, \ldots , n, 
\end{equation} 
where $ \partial_j \equiv \partial / \partial x_j$ and function $H$, which is usually taken to be a time-independent first integral, plays the role of Hamiltonian. The $J_{ij}(x)$ are the entries of an $n \times n$ structure matrix ${\cal J}$ (which can be degenerate in rank) and they have the property of being solutions of the Jacobi identities: 
\begin{equation}
     \label{plnjac}
     \sum_{l=1}^n ( J_{li} \partial_l J_{jk} + J_{lj} \partial_l J_{ki} + 
     J_{lk} \partial_l J_{ij} ) = 0 \:\; , \;\:\;\: i,j,k=1, \ldots ,n
\end{equation}
The $J_{ij}$ must also verify an additional skew-symmetry condition:
\begin{equation}
     \label{plnsksym}
     J_{ij} =  - J_{ji} \;\:\:\:\: \mbox{for all} \:\; i,j
\end{equation}

One of the reasons justifying the importance and flexibility of the Poisson representation is the (at least) local equivalence bewteen Poisson systems and classical Hamiltonian systems, as stated by Darboux Theorem \cite{olv1}. This explains that Poisson systems can be seen, to a large extent, as a generalization of classical Hamiltonian systems (for instance, allowing for odd-dimensional flows). The issue of describing a given vector field not explicitly written in the form (\ref{plnnham}) in terms of a Poisson structure is a fundamental question in this context, which still remains as an open problem \cite{9}-\cite{hyg1},\cite{5}-\cite{21}. This is a nontrivial decomposition to which important efforts have been devoted in past years in a variety of approaches. The source of the difficulty is obviously twofold: First, a known constant of motion of the system able to play the role of the Hamiltonian is required. And second, it is necessary to find a suitable structure matrix for the vector field. Consequently, finding a solution of the Jacobi identities (\ref{plnjac}) complying also with conditions 
(\ref{plnsksym}) is unavoidable. This explains the attention deserved in the literature by the obtainment and classification of skew-symmetric solutions of the Jacobi equations 
\cite{olv1}-\cite{byv2},\cite{nut1}-\cite{cyl1},\cite{5}-\cite{29}.

As far as the Jacobi identities constitute a set of 
nonlinear coupled PDEs, the characterization of $n$-dimensional solutions has 
followed, roughly speaking, a sequence of increasing nonlinearity. Thus we can speak of 
constant structure matrices (including the classical symplectic matrices), Lie-Poisson structures \cite{olv1,29}, affine-linear structures \cite{bha1}, and quadratic 
structures \cite{byv2,pla1,byr1,lyx1}, as well as some families of structure matrices which 
may contain functions of arbitrary nonlinearity \cite{byv4}. However, the set of 
solutions of system (\ref{plnjac}-\ref{plnsksym}) seems to be still mostly unexplored. 
Perhaps the only exception to this situation is that of three-dimensional (3-d in what follows) 
vector fields, which constitute an important case which has been repeatedly considered in the 
literature and is the best understood at present 
\cite{7,17},\cite{9}-\cite{hyg1},\cite{nut1,nut2,27,byv1,bs1,bs2}. In dimension three, the 
strategy for finding suitable skew-symmetric solutions of the Jacobi equations has often been 
problem-dependent. In this sense, we can find recipes based on the use of either convenient 
{\em ansatzs\/} for the solution \cite{gyn1,nut1,nut2,27}, or symmetry considerations 
\cite{7,17}, or the knowledge of additional information about the system, such as the existence of a constant of motion \cite{hyg1,byv1,8}. Additionally, in the 3-d situation it is also possible to recast the problem (\ref{plnjac}-\ref{plnsksym}) in equivalent forms which may be more suitable for the determination of the desired solutions \cite{9}-\cite{hyg1}. This is certainly a more elaborate state of affairs than the one existing in the general $n$-dimensional case. Moreover, it is worth recalling that the 3-d scenario is particularly relevant for several reasons. First, a large number of 3-d systems arising in very diverse fields have a Poisson structure \cite{7,17},\cite{9}-\cite{hyg1},\cite{nut1,nut2,27,8,bs2}. Therefore 3-d Poisson structures are the natural framework for the analysis of such systems. In second place, dimension three corresponds to the first nontrivial case where a Poisson structure does not imply a symplectic structure. In other words, it is the simplest meaningful kind of Poisson structures which is not symplectic. Finally, three is the lowest dimension for which the Jacobi identities are not always identically verified (recall that every skew-symmetric $2 \times 2$ matrix is a structure matrix). Since the complexity of equations (\ref{plnjac}-\ref{plnsksym}) is increasing with the dimension $n$, the 3-d case is the simplest nontrivial one as well as a natural first approach to the full problem of analyzing the solutions of (\ref{plnjac}-\ref{plnsksym}).

In this work a new family of skew-symmetric solutions of the three-dimensional Jacobi equations (\ref{plnjac}-\ref{plnsksym}) is considered. Such family is very general, and in particular it is  defined in terms of functions of arbitrary nonlinearity. This explains that well-known three-dimensional Poisson structures and systems now happen to appear embraced as particular cases, as it will be seen. Moreover, this unification is not only conceptual. In fact, the new familiy is amenable to explicit and detailed analysis. In particular, it is possible to explicitly determine features such as the symplectic structure and the construction of the Darboux canonical form. The advantage of these common strategies is that they are simultaneously valid for all the particular cases which can now be analyzed in a unified and more economic way, instead of using a case-by-case approach. In addition, the methods developed are valid globally in phase space, thus ameliorating the usual scope of Darboux theorem which does only guarantee, in principle, a local reduction. The possibility of constructing the Darboux canonical form is also remarkable in view that the practical determination of Darboux coordinates is a complicated task in general, which has been carried out only for a very limited sample of systems. 

The structure of the article is as follows. In Section 2 the results enumerated above, leading 
to the explicit determination and analysis of a new family of skew-symmetric solutions of the 3-d Jacobi equations, are presented. Section 3 contains several examples which illustrate the 
theory. The work is concluded in Section 4 with some final remarks.

\pagebreak
\begin{flushleft}
{\bf 2. A family of Poisson structures and its global analysis}
\end{flushleft}

The first result to be presented is the following one:

\mbox{}

\noindent{\bf Theorem 1.} 
{\em Let  $\{ \eta (x), \phi_1(x_1), \phi_2(x_2), \phi_3(x_3) \}$ be a set of functions defined in an open set $\Omega \subset I \!\! R^3$, all of which are $C^1(\Omega)$ and nonvanishing in $\Omega$. In addition, let $\kappa _{ij}$, $i,j = 1,2,3$, be arbitrary real constants that 
constitute a skew-symmetric matrix 
\begin{equation}
\label{plnkappa1}
	\kappa _{ij} + \kappa _{ji}=0 \: , \:\:\: \mbox{\rm for all} \:\:\: i,j
\end{equation}
and satisfy the zero-sum condition
\begin{equation}
\label{plnkappa2}
	\kappa _{12} + \kappa _{23} + \kappa _{31} = 0 
\end{equation}
Then ${\cal J}=(J_{ij})$ is a family 3-d Poisson structures which is globally defined in $\Omega$, with }
\begin{equation}
\label{plnsol1}
	J_{ij}(x)= \eta (x) \left( \psi_i(x_i) - \psi_j(x_j) + \kappa_{ij} \right) 
	\sum_{k=1}^3 (\epsilon _{ijk})^2 \phi_k(x_k) 
	\:\: , \:\:\:\:\: i,j = 1, 2, 3 \: ,
\end{equation}
{\em 
where $\epsilon _{ijk}$ denotes the Levi-Civita symbol, and for every $i=1,2,3$, function $\psi_i(x_i) $ denotes one of the primitive functions of $\phi_i(x_i)$. 
}

\mbox{}

\noindent{\bf Proof.} Note that for $n=3$, system (\ref{plnjac}-\ref{plnsksym}) actually consists of the following independent nonlinear equation:
\begin{equation}
\label{plnjac3df}
     J_{12} \partial_1 J_{31} - J_{31} \partial_1 J_{12} + 
     J_{23} \partial_2 J_{12} - J_{12} \partial_2 J_{23} + 
     J_{31} \partial_3 J_{23} - J_{23} \partial_3 J_{31} = 0 
\end{equation}
Consider first family (\ref{plnsol1}) in the particular case $\eta (x) =1$. For this, let $J^*_{ij}(x) \equiv J_{ij}(x)/ \eta (x)$ in (\ref{plnsol1}). Then, substitution in (\ref{plnjac3df}) produces after some algebra:
\[
   J^*_{12} \partial_1 J^*_{31} - J^*_{31} \partial_1 J^*_{12} + J^*_{23} \partial_2 J^*_{12} - 
   J^*_{12} \partial_2 J^*_{23} + J^*_{31} \partial_3 J^*_{23} - J^*_{23} \partial_3 J^*_{31} =  
	-2 \phi_1 \phi_2 \phi_3 ( \kappa_{12} + \kappa_{23} + \kappa_{31})
\]
This demonstrates the result for the case $\eta = 1$. For general $\eta$ it suffices to recall \cite{gyn1,bs2} that in the 3-d case $\eta \cdot {\cal J}$ is a structure matrix for every arbitrary nonvanishing $C^1$ function $\eta (x)$ and for every structure matrix ${\cal J}$. This completes the proof. \hfill {\Large $\Box$}

\mbox{}

Now some remarks are in order. In first place, it is useful for what is to follow to give the explicit form of the components of ${\cal J}$ for family (\ref{plnsol1}), which are:
\[
	\left\{ \begin{array}{c}
	J_{12}(x) =  \eta (x) \left( \psi_1(x_1) - \psi_2(x_2) + \kappa_{12} \right) 
		\phi_3(x_3) \\
	J_{23}(x) = \eta (x) \left( \psi_2(x_2) - \psi_3(x_3) + \kappa_{23} \right) 
		\phi_1(x_1) \\
	J_{31}(x) = \eta (x) \left( \psi_3(x_3) - \psi_1(x_1) + \kappa_{31} \right) 
		\phi_2(x_2) 
	\end{array} \right.
\]

As indicated in the theorem, for every $i$ the primitive $\psi_i(x_i)$ of $\phi_i(x_i)$ must be  chosen to be one and the same for all the entries of ${\cal J}$. However, the specific choice is actually arbitrary. To see this it suffices to notice that if a different integration constant is selected, for instance after replacing $\psi_i(x_i)$ by $\psi_i(x_i) + k_i$ for every $i$, then the outcome is also a member of the solution family, this time with constants 
$\tilde{\kappa}_{ij} = \kappa_{ij} + k_i - k_j$, which also verify (\ref{plnkappa1}-\ref{plnkappa2}). Thus conditions (\ref{plnkappa1}-\ref{plnkappa2}) express in a generalized form this degree of freedom associated with the choice of the primitives of functions $\phi_i(x_i)$.

Secondly, notice that the form of the Poisson structures we are dealing with is such that only two possibilities exist regarding the vanishing of the independent entries $(J_{12},J_{23},J_{31})$ at a given point, namely: {\em (i)\/} either none or one of them vanishes (case of rank two), or {\em (ii)\/} all of them vanish (case of zero rank). To see this, it is convenient to define the functions:
\[
	\chi_{ij}(x_i,x_j) \equiv \psi_i(x_i) - \psi_j(x_j) + \kappa_{ij} 
	\:\: , \:\:\:\:\: i,j = 1, 2, 3 
\]
Thus it is clearly not possible that only two of such entries $(J_{12},J_{23},J_{31})$ vanish at the same point, as a consequence of the zero-sum relation $\chi_{12}(x_1,x_2)+\chi_{23}(x_2,x_3)+\chi_{31}(x_3,x_1)=0$. 

To conclude, it is interesting for what is to come to recall the physical interpretation of the degree of freedom corresponding to the factor $\eta (x)$, namely the fact that in the 3-d case $\eta \cdot {\cal J}$ is a structure matrix if and only if ${\cal J}$ is \cite{gyn1,bs2}. Such result is not generally valid for dimension $n \geq 4$, as it can be easily verified. The interpretation of such three-dimensional feature is naturally associated to time reparametrizations \cite{gyn1,bs2}, which are transformations of the form
\begin{equation}
	\label{plnntt}
	\mbox{d}\tau = \frac{1}{\eta (x)}\mbox{d}t
\end{equation}
where $t$ is the initial time variable, $\tau$ is the new time and $\eta (x) : \Omega 
\rightarrow I \!\! R$ is a $C^1(\Omega)$ function which does not vanish in $\Omega$. Thus, if 
\begin{equation}
\label{pln3dpos}
	\frac{\mbox{d}x}{\mbox{d}t} = {\cal J} \cdot \nabla H
\end{equation}
is an arbitrary three-dimensional Poisson system defined in $\Omega$, then every time reparametrization (\ref{plnntt}) leads from (\ref{pln3dpos}) to the system:
\begin{equation}
	\label{pln3dposntt}
	\frac{\mbox{d}x}{\mbox{d} \tau} = \eta {\cal J} \cdot \nabla H
\end{equation}
Therefore, in the 3-d case time reparametrizations (\ref{plnntt}) preserve the Poisson structure, this time with structure matrix $\eta {\cal J}$ in (\ref{pln3dposntt}). This is not the case in general for $n \geq 4$, as mentioned.

We can now characterize some of the properties of the family identified in Theorem 1: 

\mbox{}

\noindent {\bf Theorem 2.} 
{\em Let ${\cal J}= (J_{ij})$ be a Poisson structure of the form (\ref{plnsol1}) characterized in Theorem 1, which is defined in an open domain $\Omega \subset I \!\! R^3$ and such that for a given pair $(i,j)$ one has $\chi_{ij}(x_i,x_j) \neq 0$ everywhere in $\Omega$. Then Rank(${\cal J}$)$=2$ in $\Omega$ and a Casimir invariant for ${\cal J}$ is }
\begin{equation}
\label{plncas}
	C_{k}(x) = \frac{\psi_j(x_j) - \psi_k(x_k) + \kappa_{jk}}{\psi_i(x_i) - \psi_j(x_j) 
		+ \kappa_{ij}} = \frac{\chi_{jk}(x_j,x_k)}{\chi_{ij}(x_i,x_j)}
\end{equation}
{\em 
where $(i,j,k)$ is a cyclic permutation of $(1,2,3)$. Moreover, every Casimir invariant 
(\ref{plncas}) is globally defined in $\Omega$ and belongs to $C^2(\Omega)$.
}

\mbox{}

\noindent{\bf Proof.} After some algebra it is not difficult to demonstrate that $\partial _i C_a = -( \eta \chi^2_{bc})^{-1}J_{jk}$, where both $(a,b,c)$ and $(i,j,k)$ are cyclic permutations of $(1,2,3)$. With the help of this property the result can be directly shown through the verification of the fact that ${\cal J} \cdot \nabla C_k = 0$ for each of the three cases $k=1,2,3$ indicated. The statement is completed taking into account the $C^1( \Omega )$ property of the $\phi_i(x_i)$ . \hfill {\Large $\Box$}

\mbox{}

Therefore it is possible to give the explicit list of Casimir invariants corresponding to the three complementary cases just analyzed:
\[
	C_{1}(x) = 	\frac{\psi_3(x_3) - \psi_1(x_1) + \kappa_{31}}{\psi_2(x_2) - \psi_3(x_3) + 		\kappa_{23}} = \frac{\chi_{31}(x_3,x_1)}{\chi_{23}(x_2,x_3)}
	\:\:\;\;\; \mbox{\rm if } \:\: \chi_{23}(x_2,x_3) \neq 0 \:\;\; \mbox{\rm in } \:\: 
	\Omega .
\]
\[
	C_{2}(x) = 	\frac{\psi_1(x_1) - \psi_2(x_2) + \kappa_{12}}{\psi_3(x_3) - \psi_1(x_1) + 		\kappa_{31}} = \frac{\chi_{12}(x_1,x_2)}{\chi_{31}(x_3,x_1)}
	\:\:\;\;\; \mbox{\rm if } \:\: \chi_{31}(x_3,x_1) \neq 0 \:\;\; \mbox{\rm in } \:\: 
	\Omega .
\]
\[
	C_{3}(x) =	\frac{\psi_2(x_2) - \psi_3(x_3) + \kappa_{23}}{\psi_1(x_1) - \psi_2(x_2) + 		\kappa_{12}} = \frac{\chi_{23}(x_2,x_3)}{\chi_{12}(x_1,x_2)}
	\:\:\;\;\; \mbox{\rm if } \:\: \chi_{12}(x_1,x_2) \neq 0 \:\;\; \mbox{\rm in } \:\: 
	\Omega .
\]
Notice the symmetry of such a choice, since $C_1C_2C_3=1$ when all of them are defined in $\Omega$. The previous results allow the constructive and global determination of the Darboux canonical form for this kind of Poisson structures:

\mbox{}

\noindent {\bf Theorem 3.} 
{\em Let $\Omega \subset I \!\! R^3$ be an open domain where a Poisson system (\ref{plnnham}) with $n=3$ is defined everywhere, for which ${\cal J} = (J_{ij})$ is a structure matrix of the form (\ref{plnsol1}) characterized in Theorem 1, and such that for a given pair $(i,j)$ one has $\chi_{ij}(x_i,x_j) \neq 0$ everywhere in $\Omega$. Then such Poisson system can be globally reduced in $\Omega$ to a one degree of freedom Hamiltonian system and the Darboux canonical form is accomplished globally in $\Omega$ in the new coordinate system $\{y_1,y_2,y_3\}$ and the new time $\tau$, where $\{y_1,y_2,y_3\}$ are given by the diffeomorphism globally defined in $\Omega$ 
\begin{equation}
\label{plndarbco}
	y_i (x) = x_i \;\: , \;\:\;\:
	y_j (x) = x_j \;\: , \;\:\;\:
	y_k (x) = -C_k(x) \;\: ,
\end{equation}
in which $(i,j,k)$ is a cyclic permutation of $(1,2,3)$ and $C_k(x)$ is the Casimir invariant (\ref{plncas}); while the new time $\tau$ is defined by the time reparametrization of the form (\ref{plnntt}):
}
\begin{equation}
\label{plndarbntt}
	\mbox{d} \tau = J_{ij}(x(y)) \mbox{d} t
\end{equation}

\mbox{}

\noindent{\bf Proof.} Only the case $\chi_{12}(x_1,x_2) \neq 0$ will be considered here, since the analysis of the other two cases is analogous. Note that, according to Theorem 2, the Darboux theorem is applicable because ${\cal J}$ has constant rank 2 everywhere in $\Omega$. Recall also that, after a general diffeomorphism $y = y(x)$, an arbitrary structure matrix ${\cal J}(x)$ is transformed into another one ${\cal J'}(y)$ as:
\begin{equation}
\label{plnjdiff}
      J'_{ij}(y) = \sum_{k,l=1}^n \frac{\partial y_i}{\partial x_k} J_{kl}(x) 
	\frac{\partial y_j}{\partial x_l} \;\; , \;\:\; i,j = 1, \ldots , n
\end{equation}
The reduction can be carried out in two steps. We first perform the change of variables 
(\ref{plndarbco}), which in this case is
\begin{equation}
\label{plnd12}
	y_1 = x_1 \;\: , \;\:\;\:
	y_2 = x_2 \;\: , \;\:\;\:
	y_3 = -C_3(x) 
\end{equation}
where $C_3(x)$ is given by (\ref{plncas}). For what is to come it is necessary to explicitly write the transformation inverse of (\ref{plnd12}) which is:
\begin{equation}
\label{plninvd12}
	x_1 = y_1 \; , \;\:
	x_2 = y_2 \; , \;\:
	x_3 = \zeta_3 \left( \psi_2(y_2) + \kappa_{23} + \left( \psi_1(y_1) - \psi_2(y_2) + 
		\kappa_{12} \right) y_3 \right) 
\end{equation}
where function $\zeta_3$ is the inverse function of $\psi_3(x_3)$. Note that $\zeta_3$ exists and is differentiable in $\tilde{\Omega} = \psi_3( \Omega )$. The examination of (\ref{plnd12}-\ref{plninvd12}) easily shows that the variable transformation (\ref{plnd12}) to be performed exists and is a diffeomorphism everywhere in $\Omega$ as a consequence that by hipothesis we have $\chi_{12}(x_1,x_2) \neq 0$ and $\phi _3(x_3) \neq 0$ in $\Omega$. Then, according to (\ref{plncas}) and (\ref{plnd12}), and taking (\ref{plnjdiff}) into account, after some algebra we are led to
\begin{equation}
\label{plnjdarb1}
	{\cal J'}(y) = J_{12}(x(y)) \left( \begin{array}{ccc}
	 0 & 1 & 0 \\ -1 & 0 & 0 \\ 0 & 0 & 0 \end{array} \right)
\end{equation}
where from equations (\ref{plnsol1}) and (\ref{plninvd12}) we have
\begin{equation}
\label{plnj12ntt}
	J_{12}(x(y)) = \eta (y_1,y_2,x_3(y)) \left( \psi_1(y_1) - \psi_2(y_2) + \kappa_{12} 
		\right) \phi_3(x_3(y))
\end{equation}
The explicit dependence of $x_3(y)$ is obviously the one given in (\ref{plninvd12}) and was not displayed in (\ref{plnj12ntt}) for the sake of clarity. Note that $J_{12}(x(y))$ is nonvanishing in $\Omega ' = y(\Omega)$ and $C^1(\Omega ')$. These properties allow the accomplisment of the second step of the reduction which is a reparametrization of time. Thus, making use of (\ref{plnj12ntt}) in equation (\ref{plndarbntt}), the transformation $\mbox{d} \tau = J_{12}(x(y)) \mbox{d} t$ is performed. According to (\ref{plnntt}-\ref{pln3dposntt}) this leads from the structure matrix (\ref{plnjdarb1}) to the Darboux one:
\begin{equation}
\label{plnjdarb}
	{\cal J}_D (y) = \left( \begin{array}{ccc}
	 0 & 1 & 0 \\ -1 & 0 & 0 \\ 0 & 0 & 0 \end{array} \right)
\end{equation}
The reduction is thus globally completed.  \hfill {\Large $\Box$}

\mbox{}

The previous results can be now illustrated by means of some examples. This is the aim of the next section.

\pagebreak

\begin{flushleft}
{\bf 3. Examples}
\end{flushleft}

\noindent{\bf Example 1.} Poisson structures for the Halphen equations and the system of circle maps.

\mbox{}

Let us first consider the following Poisson structure which has deserved some attention regarding the analysis of the Halphen system \cite{gyn1}:
\begin{equation}
\label{plnjhalp1}
	{\cal J} = \eta (x) \left( \begin{array}{ccc}
		    0     & x_1 - x_2 & x_1 - x_3 \\
		x_2 - x_1 &     0     & x_2 - x_3 \\
		x_3 - x_1 & x_3 - x_2 &     0 
	\end{array} \right)
\end{equation}
with 
\begin{equation}
\label{plnjhalp2}
	\eta (x) = (2(x_1-x_2)(x_2-x_3)(x_3-x_1))^{-1}
\end{equation}
It can be seen that the structure matrix (\ref{plnjhalp1}-\ref{plnjhalp2}) belongs to the family (\ref{plnsol1}) with $\psi_i(x_i)=x_i$ and $\kappa_{ij} = 0$ for all $i,j=1,2,3$, provided $x_i \neq x_j$ in $\Omega$ for every pair $i \neq j$. If this is the case, function $\eta (x)$ is $C^1 ( \Omega )$ and nonvanishing in $\Omega$. Note that this condition also implies $\chi _{ij}(x_i,x_j) \neq 0$ (and therefore $J_{ij}(x) \neq 0$) in $\Omega$ for every pair $i \neq j$. In order to perform the Darboux reduction it should be noted that every Casimir invariant (\ref{plncas}) is now defined in $\Omega$ and can thus be employed. For instance, we can focus on $C_3(x)$:
\begin{equation}
\label{plnhcas}
	C_3(x) = \frac{x_2-x_3}{x_1-x_2}
\end{equation}
Therefore the reduction to Darboux form now makes use of the following diffeomorphism
\[
	y_1 = x_1 \;\: , \;\:\;\:
	y_2 = x_2 \;\: , \;\:\;\:
	y_3 = -C_3(x) 
\]
with $C_3(x)$ given by (\ref{plnhcas}). The inverse of this transformation is then: 
\[
	x_1 = y_1 \;\: , \;\:\;\:
	x_2 = y_2 \;\: , \;\:\;\:
	x_3 = y_2 + (y_1 - y_2) y_3
\]
After applying (\ref{plnjdiff}) the outcome is that ${\cal J}$ in (\ref{plnjhalp1}-\ref{plnjhalp2}) is transformed into:
\[
	{\cal J'} = (y_1 - y_2) \eta (y_1,y_2,y_2 + y_3 (y_1 - y_2)) \left( \begin{array}{ccc}
	0 & 1 & 0 \\ -1 & 0 & 0 \\ 0 & 0 & 0 \end{array} \right) \equiv
	\tilde{J}_{12}(y) \left( \begin{array}{ccc}
	0 & 1 & 0 \\ -1 & 0 & 0 \\ 0 & 0 & 0  \end{array} \right)
\]
with $\tilde{J}_{12}(y)=(2(y_1-y_2)^2y_3(1-y_3))^{-1}$. The reduction is then completed by means of the time reparametrization $\mbox{d} \tau = \tilde{J}_{12}(y) \mbox{d} t$, which finally leads to the Darboux canonical form (\ref{plnjdarb}) with $y_3$ acting as the decoupled Casimir and $(y_1,y_2)$ as classical conjugate Hamiltonian variables. 

To conclude, it is worth mentioning in this context the Poisson structure appearing in the study of the system of circle maps \cite{gyn1}. Such structure is of the form (\ref{plnjhalp1}), but now having 
\[
	\eta (x) = -((x_1-x_2)(x_2-x_3)(x_3-x_1))^{-1}
\]
Thus the conditions for the regularity of the functions are exactly the same, the functions $\phi_i(x_i)$ and $\psi_i(x_i)$ retain their definitions, and the constants $\kappa _{ij}$ have the same zero values, than in the case of the Poisson structure for the Halphen system. The difference existing in $\eta (x)$ does not induce variations in the form of the Casimir invariants, in the diffeomorphic changes of variables leading to the Darboux reduction, or in the conditions indicating when all of them are properly defined. Consequently these results also remain valid in the context of the Poisson structure for circle maps. 

\mbox{}

\noindent{\bf Example 2.} Euler top.

\mbox{}

As a second example, the following cubic and homogeneous Poisson structure appearing \cite{gyn1} in the analysis of the Euler equations for a triaxial top will be considered:
\begin{equation}
\label{plnjtop}
	J_{12} = ( \alpha_2 x_1^2 - \alpha_1 x_2^2) x_3 \;\; , \;\;\;\; 
	J_{23} = ( \alpha_3 x_2^2 - \alpha_2 x_3^2) x_1 \;\; , \;\;\;\; 
	J_{31} = ( \alpha_1 x_3^2 - \alpha_3 x_1^2) x_2 
\end{equation}
where the $\alpha_i$ are real constants related to the principal moments of inertia $I_i$ of the top according to the expressions:
\[
	\alpha_1 = \frac{I_2-I_3}{I_2 I_3} \;\; , \;\;\;\; 
	\alpha_2 = \frac{I_3-I_1}{I_1 I_3} \;\; , \;\;\;\; 
	\alpha_3 = \frac{I_1-I_2}{I_1 I_2} 
\]
Assuming that $\alpha _1 \alpha _2 \alpha _3 \neq 0$, equations (\ref{plnjtop}) can be equivalently written as: 
\begin{equation}
\label{plnjtopf}
	\left\{ \begin{array}{c}
	J_{12}= {\displaystyle \frac{1}{2 \alpha _1 \alpha _2 \alpha _3}}
		( \alpha_2 \alpha_3 x_1^2 - \alpha_1 \alpha_3 x_2^2)(2 \alpha_1 \alpha_2 x_3) \\
	\mbox{} \\
	J_{23}= {\displaystyle \frac{1}{2 \alpha _1 \alpha _2 \alpha _3}}
		( \alpha_1 \alpha_3 x_2^2 - \alpha_1 \alpha_2 x_3^2)(2 \alpha_2 \alpha_3 x_1) \\
	\mbox{} \\
	J_{31}= {\displaystyle \frac{1}{2 \alpha _1 \alpha _2 \alpha _3}}
		( \alpha_1 \alpha_2 x_3^2 - \alpha_2 \alpha_3 x_1^2)(2 \alpha_1 \alpha_3 x_2)
	\end{array} \right.
\end{equation}
Expressed in this way, the structure matrix (\ref{plnjtopf}) can be recognized as a member of family (\ref{plnsol1}) with 
\[
\eta = (2 \alpha _1 \alpha _2 \alpha _3)^{-1} \;\; , \;\;\;\; 
\psi_1(x_1) = \alpha_2 \alpha_3 x_1^2 \;\; , \;\;\;\; 
\psi_2(x_2) = \alpha_1 \alpha_3 x_2^2 \;\; , \;\;\;\; 
\psi_3(x_3) = \alpha_1 \alpha_2 x_3^2
\]
and $\kappa_{ij} = 0$ for all $i,j=1,2,3$. Since functions $\phi_i(x_i)$ must be nonvanishing in $\Omega$, this implies that in what follows the Poisson structure (\ref{plnjtopf}) is to be analyzed in an open subset of $\{ (x_1,x_2,x_3) \in I \!\! R^3 : x_1x_2x_3 \neq 0 \}$. In addition, according to (\ref{plncas}) we can employ different forms for the Casimir invariant. For instance, if $\chi_{12}(x_1,x_2) = \alpha_2 \alpha_3 x_1^2 - \alpha_1 \alpha_3 x_2^2 \neq 0$ in $\Omega$, we have:
\begin{equation}
\label{plntopc3}
	C_{3}(x) = \frac{\alpha_1 \alpha_3 x_2^2 - \alpha_1 \alpha_2 x_3^2}{\alpha_2 \alpha_3 x_1^2 - \alpha_1 \alpha_3 x_2^2}
\end{equation}
Then, in this case a transformation leading to the Darboux canonical form is defined by 
(\ref{plnd12}) and (\ref{plntopc3}), and its inverse is a diffeomorphism in $y( \Omega )$ given by:
\[
	x_1 = y_1 \;\; , \;\;\;\; x_2 = y_2 \;\; , \;\;\;\; 
	x_3 = \sigma_3 \left( \frac{\alpha_3}{\alpha_2}y_2^2 + \left( 
	\frac{\alpha_3}{\alpha_1}y_1^2 - \frac{\alpha_3}{\alpha_2}y_2^2 \right) y_3 \right)^{1/2}
\]
where $\sigma_3 \equiv$ sign($x_3$) denotes the usual sign function, namely $(+1)$ if $x_3 > 0$ and $(-1)$ if $x_3 < 0$ (recall that $x_3 \neq 0$ in $\Omega$). The rest of the Darboux reduction does not present special features apart from the ones indicated in the proof of Theorem 3, and therefore is omitted for the sake of conciseness.

\mbox{}

\pagebreak
\begin{flushleft}
{\bf 4. Final remarks}
\end{flushleft}

The study of skew-symmetric solutions of the Jacobi equations provides an increasingly rich perspective of finite-dimensional Poisson structures. In this letter a new family of skew-symmetric solutions of the Jacobi identities has been presented and analyzed. Interestingly, the resulting solutions are not limited to a given degree of nonlinearity. This generality implies that already known Poisson structures and systems can now be seen as particular cases and become amenable to analysis in a common framework. Therefore, this unification allows the development of general methods simultaneously valid for every particular Poisson structure considered. Specifically, it is possible to determine in an algorithmic and explicit way the Casimir invariants and the Darboux canonical form. This is interesting, as far as the determination of the Darboux coordinates is in general a nontrivial task only known for a limited sample of systems. Moreover, in this work such coordinates have been charazerized globally in phase space, therefore beyond the usual scope of Darboux theorem, which only ensures a local reduction. Together with the fact that the obtainment, classification and analysis of solutions of (\ref{plnjac}-\ref{plnsksym}) are unavoidable steps in order to recast a given differential flow as a Poisson system, whenever possible, the previous results reinforce the conclusion that the direct investigation of the Jacobi equations provides fruitful perspectives for a better understanding of finite-dimensional Poisson structures, and therefore of the scope of Hamiltonian dynamics.

\pagebreak


\begin{thebibliography}{99}
   \bibitem{olv1} P. J. Olver, Applications of Lie Groups to Differential 
      Equations, 2nd Ed. (Springer-Verlag, New York, 1993).
   \bibitem{haa1} F. Haas, J. Phys. A 35 (2002) 2925.
   \bibitem{7} D. David and D. D. Holm, J. Nonlinear Sci. 2 (1992) 241.  
   \bibitem{25} G. Picard and T. W. Johnston, Phys. Rev. Lett. 48 (1982) 1610. 
   \bibitem{17} D. D. Holm and K. B. Wolf, Physica D 51 (1991)189.  
   \bibitem{dht3} D. David, D. D. Holm and M. V. Tratnik, Phys. Rep. 187  
      (1990) 281.
   \bibitem{9} J. Goedert, F. Haas, D. Hua, M. R. Feix and L. Cair\'{o}, J. Phys. A 27 
	(1994) 6495.  
   \bibitem{gyn1} H. G\"{u}mral and Y. Nutku, J. Math. Phys. 34 (1993) 5691.
   \bibitem{hyg1} F. Haas and J. Goedert, Phys. Lett. A 199 (1995) 173.
   \bibitem{byv2} B. Hern\'{a}ndez--Bermejo and V. Fair\'{e}n, J. Math. Phys. 39 (1998) 6162. 
   \bibitem{jmaa} B. Hern\'{a}ndez--Bermejo and V. Fair\'{e}n, J. Math. Anal. Appl. 256 
	(2001) 242.
   \bibitem{nut1} Y. Nutku, Phys. Lett. A 145 (1990) 27. 
   \bibitem{nut2} Y. Nutku, J. Phys. A: Math. Gen. 23 (1990) L1145.
   \bibitem{pla1} M. Plank, J. Math. Phys. 36 (1995) 3520.
   \bibitem{27} M. Plank, Nonlinearity 9 (1996) 887. 
   \bibitem{28} M. Plank, SIAM J. Appl. Math. 59 (1999) 1540.  
   \bibitem{cyl1} J. R. Cary and R. G. Littlejohn, Ann. Physics 151 (1983) 1.
   \bibitem{byv3} B. Hern\'{a}ndez--Bermejo and V. Fair\'{e}n, Phys. Lett. A 
      241 (1998) 148. 
   \bibitem{hyct} D. D. Holm, J. E. Marsden, T. Ratiu and A. Weinstein, Phys. 
      Rep. 123 (1985) 1.
   \bibitem{mac1} R. I. McLachlan, Phys. Rev. Lett. 71 (1993) 3043.
   \bibitem{mag1} F. Magri, J. Math. Phys. 19 (1978) 1156.
   \bibitem{5} G. B. Byrnes, F. A. Haggar and G. R. W. Quispel, Physica A 272 (1999) 99. 
   \bibitem{byv1} B. Hern\'{a}ndez--Bermejo and V. Fair\'{e}n, Phys. Lett. A 234 (1997) 35. 
   \bibitem{16} S. A. Hojman, J. Phys. A 29 (1996) 667.
   \bibitem{21} C. A. Lucey and E. T. Newman, J. Math. Phys. 29 (1988) 2430. 
   \bibitem{bha1} K. H. Bhaskara, Proc. Indian Acad. Sci. Math. Sci. 100 
      (1990) 189.
   \bibitem{byr1} K. H. Bhaskara and K. Rama, J. Math. Phys. 32 (1991) 2319.
   \bibitem{8} P. Gao, Phys. Lett. A 273 (2000) 85. 
   \bibitem{byv4} B. Hern\'{a}ndez--Bermejo and V. Fair\'{e}n, Phys. Lett. A 271 (2000) 258. 
   \bibitem{bs1} B. Hern\'{a}ndez--Bermejo, Phys. Lett. A 287 (2001) 371. 
   \bibitem{bs2} B. Hern\'{a}ndez--Bermejo, J. Math. Phys. 42 (2001) 4984. 
   \bibitem{lyx1} Z.-J. Liu and P. Xu, Lett. Math. Phys. 26 (1992) 33.
   \bibitem{29} J.-L. Thiffeault and P. J. Morrison, Physica D 136 (2000) 205. 
\end{thebibliography}
\end{document}